\pdfoutput=1
\documentclass[sigconf, 9pt]{acmart}
\AtBeginDocument{%
  }

\usepackage{amsmath,amssymb,amsfonts}
\usepackage{graphicx}
\usepackage{textcomp}
\usepackage{xcolor}
\usepackage{algpseudocodex}
\usepackage[ruled, linesnumbered]{algorithm2e}

\usepackage{amsthm}
\usepackage{multirow}
\usepackage{float}
\usepackage{makecell}
\usepackage{hyperref}
\usepackage{cleveref}
\usepackage{tikz}
\usepackage{commath} 
\usepackage{subfigure}
\usepackage{siunitx}
\usetikzlibrary{fit,calc}

\usepackage{multirow}
\usepackage{bm}
\newcommand{\tabincell}[2]{\begin{tabular}{@{}#1@{}}#2\end{tabular}}
\usepackage{comment}

\let\oldnl\nl
\newcommand{\nonl}{\renewcommand{\nl}{\let\nl\oldnl}}

\newtheorem{problem}{Problem}

\newtheorem{theorem}{Theorem}
\newtheorem{proposition}{Proposition}

\usepackage[braket, qm]{qcircuit}

\usepackage{amssymb}

\usepackage{subcaption}

\usepackage{enumitem}

\usepackage[flushleft]{threeparttable}

\renewcommand{\ket}[1]{ \left| #1 \right \rangle}

\copyrightyear{2024}
\acmYear{2024}
\setcopyright{acmlicensed}\acmConference[DAC '24]{61st ACM/IEEE Design Automation Conference}{June 23--27, 2024}{San Francisco, CA, USA}
\acmBooktitle{61st ACM/IEEE Design Automation Conference (DAC '24), June 23--27, 2024, San Francisco, CA, USA}
\acmDOI{10.1145/3649329.3657312}
\acmISBN{979-8-4007-0601-1/24/06}

\begin{document}

\SetKwIF{If}{ElseIf}{Else}{if}{}{else if}{else}{}%
\SetKwFor{While}{while}{}{}%
\SetKwFor{For}{for}{}{}%

\title{Boolean Matching Reversible Circuits:
Algorithm and Complexity
}

\author{Tian-Fu Chen$^{1,4,5}$ and Jie-Hong R. Jiang$^{1,2,3,4,5}$}
\affiliation{
  \institution{$^1$\textit{Graduate School of Advanced Technology, National Taiwan University}, Taipei, Taiwan}
  \institution{$^2$\textit{Graduate Institute of Electronics Engineering, National Taiwan University}, Taipei, Taiwan}
  \institution{$^3$\textit{Department of Electrical Engineering, National Taiwan University}, Taipei, Taiwan}
  \institution{$^4$\textit{Center for Quantum Science and Engineering, National Taiwan University}, Taipei, Taiwan}
  \institution{$^5$\textit{Physics Division, National Center for Theoretical Sciences}, Taipei, Taiwan}
  \institution{\{\,d11k42001, jhjiang\,\}\,@ntu.edu.tw}
  \country{}
}

\begin{abstract}
Boolean matching is an important problem in logic synthesis and verification.
Despite being well-studied for conventional Boolean circuits, its treatment for reversible logic circuits remains largely, if not completely, missing.
This work provides the first such study.
Given two (black-box) reversible logic circuits that are promised to be matchable, we check their equivalences under various input/output negation %
and permutation %
conditions subject to the availability/unavailability of their inverse circuits.
Notably, among other results, we show that the equivalence up to input negation and permutation is solvable in quantum polynomial time, while its classical complexity is exponential. 
This result is arguably the first demonstration of quantum exponential speedup in solving design automation problems.
Also, as a negative result, we show that the equivalence up to both input and output negations is not solvable in quantum polynomial time unless UNIQUE-SAT is, which is unlikely.
This work paves the theoretical foundation of Boolean matching reversible circuits for potential applications, e.g., in quantum circuit synthesis.

\end{abstract}


\settopmatter{printacmref=false}

\maketitle

\section{Introduction}
\label{sec:introduction}
\emph{Reversible logic circuits}, simply called \emph{reversible circuits}, are a class of Boolean circuits with $n$ inputs and $n$ outputs whose functions $\mathbb{B}^n \rightarrow \mathbb{B}^n$ form a bijection (one-to-one and onto) mapping, i.e., a permutation.
Fundamentally, logical reversibility prevents information erasure and is a necessary condition for computation with extremely low energy dissipation \cite{heat,reversible_original}.
Remarkably, quantum computing \cite{textbook} is intrinsically reversible. 
Every quantum gate, i.e., unitary operator, is reversible.\footnote{Although every quantum circuit is reversible, not every quantum circuit is referred to as a reversible circuit unless its unitary operator is a permutation matrix.}
Reversible circuit synthesis has been extensively studied, e.g., 
\cite{transform,reversible, soeken2017hierarchical, soeken2012revkit, soeken2015embedding},
largely motivated by quantum computation due to the fact that several important quantum algorithms, such as Grover's algorithm \cite{grover}, Simon's algorithm \cite{simon}, etc., have oracle circuits, which are reversible circuits, as a key building block.
In addition, reversible circuits also have applications in signal processing, cryptography, computer graphics, and nanotechnologies \cite{reversible}.
Constructing optimal reversible circuits is of practical importance.
In this work, we are concerned with \emph{Boolean matching} of reversible circuits.

Conventionally, Boolean matching checks whether two (irreversible) logic circuits are equivalent up to the negation and/or permutation of inputs and/or outputs \cite{BooM,LSBM}.
It plays a vital role in logic synthesis and verification tasks such as technology mapping \cite{benini1997survey}, %
engineering change order (ECO) \cite{ECO}, etc.
Although well-studied for conventional Boolean circuits, Boolean matching on reversible circuits is largely, if not completely, missing.
This work provides the first comprehensive study of Boolean matching reversible circuits and paves the theoretical foundation for potential applications. 
For example, template-based reversible logic synthesis \cite{transform} will greatly benefit from Boolean matching (extending from structural to functional matching), and so will oracle circuit synthesis and verification for quantum computation.

\begin{figure}[t]
    \centering
    \includegraphics[width=\columnwidth]{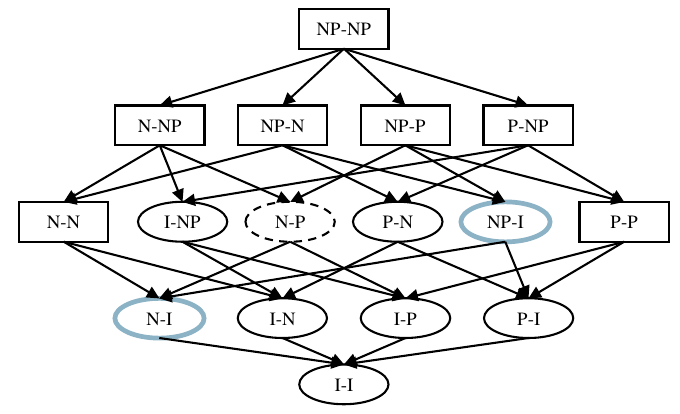}
    \caption{Domination relation among and computational complexities of various Boolean matching equivalences.}
    \label{fig:dominance}
\end{figure}

Given two (black-box) reversible logic circuits, also referred to as \emph{oracles}, that are promised to be matchable, we explore the algorithm and complexity for Boolean matching on reversible circuits under different equivalences up to negation and/or permutation on their inputs and/or outputs.
We define ``X-Y equivalence'' for X, Y $\in \{$I, N, P, NP$\}$, where X and Y denote equivalence conditions on the input and output sides, respectively, of the circuits, and I, N, P, and NP stand for identity, negation equivalence, permutation equivalence, and negation plus permutation equivalence, respectively. 
In total, there can be 16 combinations of equivalences,
whose dominance relations are shown in the graph of \Cref{fig:dominance}, where an edge $(A,B)$ from $A$ to $B$ indicates equivalence $B$ is subsumed by $A$.
Note that the I-I equivalence is just the typical combinational equivalence problem without needing to identify negation and permutation conditions.

This work contributes to the first comprehensive characterization of the computational complexity in terms of the number of oracle access required to compute the negation/permutation conditions for the 16 equivalences.
\Cref{fig:dominance} summarizes the results. 
The equivalences in ovals and rectangles correspond to easy (solvable in classical or quantum polynomial time) and hard (no easier than UNIQUE-SAT \cite{promise}) cases, respectively. 
As positive results, we provide concrete polynomial-time classical or quantum algorithms for checking the tractable equivalences. 
We note that the two gray-blue 
colored ovals (for N-I and NP-I equivalences) indicate their quantum, but not classical, polynomial-time solvability.
To solve N-I and NP-I equivalences, we develop a quantum algorithm utilizing the \emph{swap test} \cite{swaptest} as a subroutine.\footnote{Besides the swap-test-based algorithm, we develop two more algorithms inspired by Simon's algorithm \cite{simon} and have to omit them due to space limit.}
The new algorithm achieves an exponential speedup over classical computation, adding to the rare quantum algorithm examples of attaining exponential speedup.
Arguably, it is the first quantum algorithm with an exponential speedup in solving design automation and quantum program compilation problems.
On the other hand, the dashed oval (for N-P equivalence) in \Cref{fig:dominance} indicates conditional classical polynomial-time solvability but with unknown quantum complexity.
As negative results, we prove that N-N and P-P equivalences cannot be solved in quantum polynomial time unless UNIQUE-SAT can.
Consequently, all other equivalences that dominate N-N or P-P equivalences are UNIQUE-SAT hard.

The rest of this paper is organized as follows.
\Cref{sec:preliminaries} provides essential backgrounds on reversible circuits and quantum computation.
\Cref{sec:problem} formulates the Boolean matching problem of reversible circuits.
For the equivalences that are polynomially solvable, \Cref{sec:methods} presents the corresponding algorithms and analyzes their complexities.
\Cref{sec:discussions} studies the equivalences that we prove to be harder than UNIQUE-SAT.
Finally, we conclude this work in \Cref{sec:conclusions}.

\section{Preliminaries}
\label{sec:preliminaries}

\subsection{Reversible Circuits}
\begin{figure}[t] 
\centering
\[
   \Qcircuit @C=.5em @R=.6em @!R { 
        \lstick{i_0} & \ctrl{1}  & \qw   & \qw & \rstick{o_0 = {i_0}}  \\
        \lstick{i_1} & \ctrlo{1} & \targ & \qw & \rstick{o_1 = \overline{i_1}}  \\
        \lstick{i_2} & \targ     & \qw   & \qw & \rstick{o_2 = i_2 \,\oplus\; i_0\overline{i_1}  }  \\
    } 
\]
\caption{A reversible circuit example.}    
\label{fig:reversible}     
\end{figure}
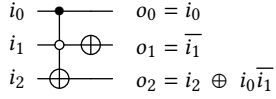

An $n$-bit reversible circuit contains $n$ input bits and $n$ output bits (depicted with $n$ lines with inputs at the left ends and outputs at the right ends) and implements a one-to-one and onto function $f:\mathbb{B}^n \rightarrow \mathbb{B}^n$, i.e., a permutation mapping.
Clearly, $f$ is reversible as its inverse function $f^{-1}$ always exists.
A reversible circuit can be generally represented by a circuit composed of multiple-controlled Toffoli (MCT) gates \cite{reversible}. 
An MCT gate has $k=0,1,2,\ldots$ control bits, each of which can be of a positive (indicated with a solid dot) or negative polarity (indicated with an empty circle), and conditionally flips a target bit (indicated with sign $\oplus$) when and only when all the negative and positive control bits are of values 0 and 1, respectively.
In the special case of $k = 0$ and $1$, the MCT gate corresponds to the \textsc{not}- and \textsc{cnot}-gate (for a positive control bit), respectively.
\Cref{fig:reversible} shows an example.

\subsection{Quantum Computation}
A \emph{quantum bit} (\emph{qubit}) can be 
in a superposition state over states $\ket{0}$ and $\ket{1}$,  generally specified, in the Dirac notation, by 
$\ket{\psi} = \alpha \ket{0} + \beta \ket{1}$, where $\alpha$ and $\beta \in \mathbb{C}$ are referred to as the \emph{probability amplitudes}, with $\abs{\alpha}^2 + \abs{\beta}^2 = 1$. 
Specifically, $\abs{\alpha}^2$ and $\abs{\beta}^2$ correspond to the probabilities for measuring the state $\ket{\psi}$ be in $\ket{0}$ and $\ket{1}$, respectively.
A quantum system of $n$ qubits can be specified by $\ket{\psi} = a_{0 \ldots 0}\ket{0 \ldots 0} + \ldots + a_{1 \ldots 1}\ket{1 \ldots 1}$, or alternatively written as a column vector $\ket{\psi} = [ a_{0 \ldots 0}, \ldots,  a_{1 \ldots 1}]^\text{T}$.
The evolution of a quantum system of $n$ qubits is governed by a sequence of quantum gates $U_1, \ldots, U_k$, for each $U_i$ being a $2^n \times 2^n$ unitary matrix (or operator) over complex numbers satisfying $U_i^\dagger=U_i^{-1}$, i.e., its conjugate transpose equals its inverse. 
The gate sequence constitutes a quantum circuit with a global unitary matrix $U$ equal to the matrix product $U_k \cdots U_1$.
Applying $U$ on an initial quantum state $\ket{\psi}$ leads to the final state $\ket{\psi'} = U\ket{\psi}$.
Given two arbitrary states $\ket{\psi_1}$ and $\ket{\psi_2}$, a unitary matrix $U$ preserves their inner product, i.e., 
$\langle \psi_1 | U^\dagger  U | \psi_2\rangle = \langle \psi_1 | \psi_2 \rangle$,
where $\langle \psi_1 | \psi_2 \rangle$ denotes the inner product of $\ket{\psi_1}$ and $\ket{\psi_2}$, and $\langle\psi_1 | U^\dagger  U | \psi_2\rangle$ is the inner product of $U\ket{\psi_1}$ and $U\ket{\psi_2}$.
Note that the function mapping defined by a reversible circuit can be represented as a permutation matrix, which is a unitary operator.

\begin{figure}[t] 
\centering
\[
   \Qcircuit @C=0.25cm @R=.5cm {
    \lstick{\ket{0}}        & \qw       & \gate{H}  & \qw & \ctrl{2} & \qw & \gate{H} & \meter & \rstick{z} \qw \\
    \lstick{\ket{\psi_1}}   & {/^n}\qw  & \qw       & \qw & \qswap   & \qw & \qw      & \qw    & \qw  \\
    \lstick{\ket{\psi_2}}   & {/^n}\qw  & \qw       & \qw & \qswap   & \qw & \qw      & \qw    & \qw  \\
    }
\]
\caption{The circuit of quantum swap test.}    
\label{fig:swaptest}     
\end{figure}
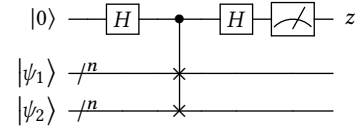

In this work, we utilize the \textit{swap test} \cite{swaptest}, a quantum computation technique used to check how much two quantum states differ, 
in some of our algorithms.
Given two $n$-qubit quantum states $\ket{\psi_1}$ and $\ket{\psi_2}$, the swap test applies the circuit shown in \Cref{fig:swaptest} and measures the first qubit.
After measurement, $z$ is either in $\ket{0}$ of probability $\frac{1}{2} + \frac{1}{2}\abs{\langle\psi_1 | \psi_2\rangle}^2$ or in $\ket{1}$ of probability $\frac{1}{2} - \frac{1}{2}\abs{\langle\psi_1 | \psi_2\rangle}^2$.
Hence, if 
$\ket{\psi_1}$ and $\ket{\psi_2}$ are identical,
i.e., $\abs{\langle\psi_1 | \psi_2\rangle} = 1$,
then the outcome $z$ is always in $\ket{0}$.
At the other extreme, if 
$\ket{\psi_1}$ and $\ket{\psi_2}$ are orthogonal, 
i.e, $\abs{\langle\psi_1 | \psi_2\rangle} = 0$,
then the outcome $z$ is in $\ket{0}$ and $\ket{1}$ with equal probability.

\section{Problem Formulation} \label{sec:problem}
The Boolean matching problem for reversible circuits can be stated as follows.

\begin{problem}[Boolean matching reversible circuits]
    Given two $n$-bit black-box reversible circuits $C_1$ and $C_2$ with $C_1$ and $C_2$ being promised to be X-Y equivalent for X, Y $\in \{$I, N, P, NP$\}$,
    we are asked to find the corresponding negation functions $\nu_x, \nu_y: \{1, \ldots, n\} \rightarrow \mathbb{B}$ and permutation functions $\pi_x, \pi_y: \{1, \ldots, n\} \rightarrow \{1, \ldots, n\}$ of X and Y that make $C_1$ and $C_2$ X-Y equivalent, where $\nu(i) = 1$ indicates the $i^\mathrm{th}$ bit being negated and $\pi(i) = j$ indicates the $i^\mathrm{th}$ bit being permuted to the $j^\mathrm{th}$ bit.
\label{problem:problem}    
\end{problem}  
In the sequel, we omit the subscripts $x,y$ in $\nu$ and $\pi$ when the input/output information is clear from the context or immaterial to the discussion.
We denote the reversible circuits of functions $\nu$ and $\pi$ as $C_\nu$ and $C_\pi$, respectively.
Furthermore, we do not distinguish a reversible circuit $C$ from its underlying unitary matrix.
Hence, concatenating a reversible circuit $C_A$ (on the left) followed by circuit $C_B$ (on the right) is represented as the matrix product $C_B C_A$.
For example, N-P equivalent $C_1$ and $C_2$ means there exist $C_{\nu}$ and $C_{\pi}$ to make $C_1 = C_{\pi}C_2C_{\nu}$ for some input negation function $\nu$ and output permutation function $\pi$.

Notice that \Cref{problem:problem} is a promise problem; that is, the circuits under test are promised to be Boolean matchable under certain equivalences.
Finding a solution to the promise problem is crucial to even answering the non-promise version of the problem.
It is because as long as we can solve \Cref{problem:problem}, a solution of the negation and permutation conditions can be obtained even if the equivalence relation is not promised.
With the found negation and permutation conditions, only a single round of equivalence checking is needed to validate the equivalence relation.
In contrast, without knowing the negation and permutation conditions, one may need to try an exponential number of equivalence checking rounds for all $\nu$ and $\pi$ options.

Also note that \Cref{problem:problem} assumes $C_1$ and $C_2$ are given as black boxes (oracles).
The computational complexity of finding $\nu$ and $\pi$ functions is measured in terms of the number of queries to the oracles.
A variant problem of \Cref{problem:problem} is to relax the assumption given not only $C_1$ and $C_2$ but also their inverse circuits $C_1^{-1}$ and $C_2^{-1}$.
Surely, if $C_1$ and $C_2$ are given as white boxes, we can always derive their inverse circuits as they are reversible.
In \Cref{sec:methods}, we will discuss how this relaxation may possibly simplify the computation.

\section{Tractable Equivalences}
\label{sec:methods}
In this section, we investigate Boolean matching equivalences $\{$I-NP, N-P, P-N, NP-I, N-I, I-N, I-P, P-I$\}$ that permit polynomial time identification of negation and permutation functions.
For these equivalences, we develop efficient algorithms and analyze their complexities as summarized in Table~\ref{table:compare}, where $n$ is the bit size of the circuits and $\epsilon$ is the failure probability for randomized algorithms.

\begin{table}[b]
\caption{Complexity of computing various equivalences.}
\label{table:compare}
\centering
\vspace{-0.5em}
\renewcommand\arraystretch{1.2}
\begin{threeparttable}
\begin{tabular}{|c|c|c|c|}
\hline
\multirow{2}{*}{\tabincell{c}{Inverse\\ circuit}} & \multirow{2}{*}{\tabincell{c}{Equivalence \\type}} & \multirow{2}{*}{\tabincell{c}{Computing\\paradigm}}   & \multirow{2}{*}{\tabincell{c}{Complexity}} \\
  &&& \\ 
\hline \hline
\cline{2-4}
\hline
\multirow{4}{*}{available}    & N-I*, I-N*            & classical      & $O(1)$ \\
\cline{2-4}
                                                                        & \multirow{3}{*}{\tabincell{c}{I-P*, P-I*,\\ N-P**,  P-N*,\\  I-NP*, NP-I*}} & \multirow{3}{*}{\tabincell{c}{classical}}      & \multirow{3}{*}{\tabincell{c}{$O(\log{n})$}} \\
&&&       \\ 
&&&       \\ 
\cline{2-4}
\hline
\multirow{5}{*}{\tabincell{c}{not\\ available}} & I-N                 & classical      & $O(1)$ \\
\cline{2-4}
                                                                        & I-P, I-NP           & classical      & $O(\log{n} + \log{(1/\epsilon}))$ \\
\cline{2-4}
\cline{2-4}                                                                        
                                                                        & P-I, P-N            & classical      & $O(n)$ \\
\cline{2-4}
                                                                        & N-I                 & quantum        & $O(n\log{(1/\epsilon)})$ \\
\cline{2-4}

                                                                        & NP-I                & quantum        & $O(n^2\log{(1/\epsilon)})$ \\
\hline
\end{tabular}
\begin{tablenotes}
\item
    ``*'' indicates one inverse circuit of either $C_1$ or $C_2$ is required.
\item
    ``**'' indicates both inverse circuits of $C_1$ and $C_2$ are needed.
\end{tablenotes}
\end{threeparttable}
\end{table}

\subsection{I-N Equivalence}
\label{subsec:I-N}
\begin{proposition}
For I-N equivalence, finding $\nu$ for $C_1 = C_\nu C_2$ is of $O(1)$ query complexity.
\end{proposition}
We set all inputs of $C_1$ and $C_2$ to 0 and bit-wisely compare their outputs.
Then, for $i=1,\ldots,n$, we have $\nu(i) = 1$ if and only if the output patterns of $C_1$ and $C_2$ differ at the $i^\mathrm{th}$ bit.
The computation requires only one oracle query.

\subsection{I-P Equivalence}
\label{subsec:I-P}
\begin{proposition}
For I-P equivalence, finding $\pi$ for $C_1 = C_\pi C_2$ is of $O(\log(n))$ query complexity if $C_1^{-1}$ or $C_2^{-1}$ is available, and of query complexity $O(\log(n) + \log(1/\epsilon))$ by a randomized algorithm with success probability $1-\epsilon$ if $C_1^{-1}$ and $C_2^{-1}$ are unavailable.
\end{proposition}

If $C_2^{-1}$ is available, let $C = C_1 C_2^{-1}$ by concatenating $C_2^{-1}$ and $C_1$, and thus $C$ is functionally equivalent to $C_\pi$.
We use $\lceil \log_2{n} \rceil$ input patterns to decide $\pi$ as follows.
For the $i^\mathrm{th}$ input pattern, the input of the $j^\mathrm{th}$ bit of the circuit is the $i^\mathrm{th}$ least significant bit of the binary code of index $j$.
In other words, the input sequence to the $j^\mathrm{th}$ bit of the circuit is the binary code of $j$ with the least significant bit arriving first.
Then $\pi(p) = q$ if and only if the sequential input of the $p^\mathrm{th}$ bit is the same as the sequential output of the $q^\mathrm{th}$ bit.
Hence, $O(\log{n})$ oracle queries of $C$ are needed. 
Similarly, if $C_1^{-1}$ is available, we let $C = C_2 C_1^{-1}$, which equals $C_{\pi^{-1}}$.
The same way applies to find $\pi^{-1}$, and thus $\pi$.

If both $C_1^{-1}$ and $C_2^{-1}$ are unavailable, the following randomized algorithm is applied to find $\pi$ in two steps.
First, repeat $k$ times to feed random input patterns to both $C_1$ and $C_2$ and record their output patterns, where $k$ is a number related to the failure probability to be explained shortly.
Second, for each output bit $b_1 = 1, \ldots, n$ of $C_1$, find the unique output bit $b_2$ of $C_2$ such that they share the same output sequence,
which indicates $\pi(b_1) = b_2$. 

To ensure the second step finds a unique $b_2$ for each $b_1$, there cannot be two bits sharing the same output sequence.
Hence, we require $k$ iterations of the first step to make the output sequence long enough.
Note that the range under the mapping of a reversible circuit must cover all $2^n$ combinations, and thus, different input patterns must yield different output patterns.
As long as the input patterns are uniformly generated at random, 
the output patterns are also uniformly at random.
Let $B_1(j)$ be the output sequence of the $j^\mathrm{th}$ bit in $C_1$.
Since there are $2^k$ possible output sequences for any output bit $j$, the success probability $\Pr$ that $B_1(j_1) \neq B_1(j_2) \;\forall j_1 \neq j_2$ is 
\begin{equation}
\begin{aligned}
    \Pr &= \frac{{(2^k)(2^k-1)\ldots(2^k-n+1)}}{{(2^k)}^n} 
                \geq \left(\frac{2^k-n+1}{2^k}\right)^n \\
                &\approx 1- \frac{n(n-1)}{2^k} \;\;(\text{when } n \ll 2^k).          
\end{aligned}                
\end{equation}
To make $\Pr \geq 1 - \epsilon$, we need $k \geq \log_2{\frac{n(n-1)}{\epsilon}} = O(\log(n) + \log(1/\epsilon))$.
Therefore, the query complexity of the algorithm is $O(\log(n) + \log(1/\epsilon))$.

\subsection{I-NP Equivalence}
\label{subsec:I-NP}
\begin{figure}[t]
\centering
\[
    \Qcircuit @C=0.25cm @R=.25cm {
       & \qw & \qw & \qw & \link{2}{-1}  & \qw   & \qw  &&   &&& \targ & \qw & \link{2}{-1}  & \qw  & \qw & \qw &  \\
       &     &     &     &               &       &      && = &&&       &     &               &      &     &     &  \\      
       & \qw & \qw & \qw & \link{-2}{-1} & \targ & \qw  &&   &&& \qw   & \qw & \link{-2}{-1} & \qw  & \qw & \qw & 
    }
\]
\caption{Exchanging the order between negation and permutation circuits.} 
\label{fig:swap}
\end{figure}
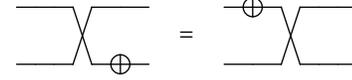

\begin{proposition}
For I-NP equivalence, finding $\nu$ and $\pi$ for $C_1 = C_\pi C_\nu C_2$ is of $O(\log(n))$ query complexity if either $C_1^{-1}$ or $C_2^{-1}$ is available, and of  complexity $O(\log(n) + \log(1/\epsilon))$ by a randomized algorithm with success probability $1 - \epsilon$ if $C_1^{-1}$ and $C_2^{-1}$ are unavailable.
\end{proposition}
Observe that the order of $C_\nu$ and $C_\pi$ can be exchanged according to the identity shown in \Cref{fig:swap}.
If $C_2^{-1}$ is available, let $C = C_1 C_2^{-1}$ by concatenating $C_2^{-1}$ and $C_1$, which is functionally equivalent to $C_\pi C_\nu$.
We temporally let $C_\pi C_\nu =  C_{\nu'} C_{\pi'}$.
To decide $\nu'$, we set all inputs to 0. 
Since all inputs have the same value, the initial permutation circuit $C_{\pi'}$ has no effect, and thus, $\nu'(i) = 1$ if and only if the output of the $i^\mathrm{th}$ bit is 1.
This step requires only one oracle query.
After $\nu'(i)$ is decided, the method mentioned in \Cref{subsec:I-P} can be slightly modified to find $\pi'$, which needs $O(\log(n))$ oracle queries.
Finally, $\nu'$ and $\pi'$ can be transformed back to get $\nu$ and $\pi$. 
Similarly, if $C_1^{-1}$ is available, we let $C =  C_2 C_1^{-1}$ and follow a similar procedure to obtain $\nu^{-1}$ and $\pi^{-1}$, and transform them back to $\nu$ and $\pi$. 

If both $C_1^{-1}$ and $C_2^{-1}$ are unavailable, a randomized algorithm similar to the one in \Cref{subsec:I-P} can be employed to find $\pi$.
The modification is that when we look for the unique $b_2$ in $C_2$ for each bit $b_1$ in $C_1$, we now recognize two kinds of $b_2$.
One is the same as before, where the $b_1^\mathrm{th}$ bit in $C_1$ has the same output sequence as the $b_2^\mathrm{th}$ bit in $C_2$, which indicates $\pi(b_1) = b_2$ and $\nu(b_2) = 0$. 
The other is that the $b_1^\mathrm{th}$ bit in $C_1$ has exactly the bitwise-flipped output sequence to the $b_2^\mathrm{th}$ bit in $C_2$, which indicates $\pi(b_1) = b_2$ and $\nu(b_2) = 1$. 
An analysis similar to that in \Cref{subsec:I-P} gives the $O(\log(n) + \log(1/\epsilon))$ complexity.

\subsection{P-I Equivalence}
\label{subsec:P-I}
\begin{proposition}
For P-I equivalence, finding $\pi$ for $C_1 = C_2 C_\pi$ is of $O(\log(n))$ query complexity if $C_1^{-1}$ or $C_2^{-1}$ is available, and of $O(n)$ complexity if $C_1^{-1}$ and $C_2^{-1}$ are unavailable.
\end{proposition}
If $C_1^{-1}$ (resp. $C_2^{-1}$) is available, we obtain $C = C_1^{-1} C_2$ (resp. $C =  C_1 C_2^{-1}$), which is functionally equivalent to $C_{\pi^{-1}}$ (resp. $C_\pi$).
As proposed in \Cref{subsec:I-P}, $\pi$ can be decided in $O(\log(n))$ number of oracle queries.

If both $C_1^{-1}$ and $C_2^{-1}$ are unavailable, $\pi$ can be found in $O(n)$ number of oracle queries with the following algorithm in two steps.
First, prepare $n$ one-hot input patterns.
The $i^\mathrm{th}$ pattern assigns 0 to all bits except one 1 to the $i^\mathrm{th}$ bit.
For the $i^\mathrm{th}$ input pattern, collect the corresponding output patterns $P_{o1,i}$ of $C_1$ and $P_{o2,i}$ of $C_2$.
Let $M_1[P_{o1,i}] = i$ and $M_2[i] = P_{o2,i}$.
Second, for each $i = 1, \ldots, n$, let
$\pi(i) = M_1[M_2[i]]$.
Thereby, $\pi$ is found in $O(n)$ oracle queries.

\subsection{N-I Equivalence}
\label{subsec:N-I}
\begin{proposition}
For N-I equivalence, finding $\nu$ for $C_1 = C_2 C_\nu$ is of $O(\log(n))$ query complexity if either $C_1^{-1}$ or $C_2^{-1}$ is available, and of $O(n \log(1/\epsilon))$ query complexity by a randomized quantum algorithm with success probability $1-\epsilon$ if $C_1^{-1}$ and $C_2^{-1}$ are unavailable but $C_1$ and $C_2$ can take quantum states as inputs.
\end{proposition}

If $C_2^{-1}$ is available, let $C = C_2^{-1} C_1 = C_\nu$.
By setting all inputs of $C$ to 0, $\nu(i) = 1$ if and only if the $i^\mathrm{th}$ bit at output is 1.
Similarly, if $C_1^{-1}$ is available, let $C = C_1^{-1} C_2 = C_{\nu^{-1}}$.
Observing $C_{\nu^{-1}}$ =  $C_{\nu}$, the same approach can be applied to decide $\nu$.

If $C_1^{-1}$ and $C_2^{-1}$ are both unavailable, \Cref{thm:N-I} provides the complexity lower bound of classical algorithms.
\vspace{-.5em}
\begin{theorem} \label{thm:N-I}
If $C_1^{-1}$ and $C_2^{-1}$ are unavailable, a classical algorithm to find the negation function for N-I equivalence needs at least $\Omega(2^{n/2})$ oracle queries.
\end{theorem} \vspace{-1.5em}
\begin{proof} \label{proof:N-I}
Given black boxes $C_1$ and $C_2$, they can be accessed only by queries, specifically, one output response per input query.
Only when $C_1$ and $C_2$ yield the same output upon two input patterns to $C_1$ and $C_2$, we can infer $\nu$ by comparing these two input patterns.
A classical algorithm can only randomly try different input patterns until $C_1$ and $C_2$ produce the same output pattern, i.e., a collision happens.
Assuming $k$ queries to $C_2$ are made to match an output pattern of $C_1$, then the probability of collisions not happening is 
\begin{equation}
\begin{aligned}
    \Pr &= \frac{{(2^n)(2^n-1)\ldots(2^n-k+1)}}{{(2^n)}^k} 
                \geq \left(\frac{2^n-k+1}{2^n}\right)^k \\
                &\approx 1- \frac{k(k-1)}{2^n} \;\;(\text{when } k \ll 2^n).       
\end{aligned}                
\end{equation}
For $\Pr$ being smaller than some constant threshold $c$, the query number $k$ should be at least $\sqrt{(1-c) \cdot 2^n}$, which yields the lower bound $\Omega(2^{n/2})$.
\end{proof}

Fortunately, assuming that the reversible circuits $C_1$ and $C_2$ can take quantum states as input, finding $\nu$ for N-I equivalence without $C_1^{-1}$ and $C_2^{-1}$ is solvable in quantum polynomial time by \Cref{alg:input-negation}.
The key idea is to utilize superposition states to disable \textsc{not}-gates, such that only one \textsc{not}-gate is active at a time.
In addition, the swap test \cite{swaptest} is exploited as a subroutine to compare two quantum states.
\Cref{alg:input-negation} proceeds in $n$ iterations, where the $i^\mathrm{th}$ iteration decides the value of $\nu(i)$.
In the $i^\mathrm{th}$ iteration, the inputs of the $i^\mathrm{th}$ qubit of both circuits are set to $\ket{0}$, while all other qubits are initialized to $\ket{+}=\frac{1}{\sqrt{2}}(\ket{0}+\ket{1})$ in lines~2 to 3.
Line~4 sets $\nu(i) = 0$ in default.
In lines~5 to 8, $k$ iterations of swap tests are performed for $k$ being determined by the failure probability as to be explained.
In line~6, we execute $C_1$ and $C_2$ on input pattern $P_{in(i)}$ to obtain their final states $C_1(P_{in(i)})$ and $C_2(P_{in(i)})$, and these two final states are sent to the swap test.
If the measurement outcome of the swap test is 1, we conclude that $\nu(i) = 1$, and the iteration is terminated, as shown in lines~7 to 8.
Otherwise, if the measurement outcome is 0 for $k$ times, then we have high confidence that $\nu(i) = 0$.

\begin{algorithm}[t]
\caption{Quantum Algorithm for N-I Equivalence}
\label{alg:input-negation}   
    \DontPrintSemicolon
    \SetArgSty{text}
    \SetDataSty{text}
    \SetKwInOut{Input}{Input}
    \SetKwInOut{Output}{Output} 
    \Input{$C_1, C_2$: Two $n$-bit reversible circuits}
    \Output{$\nu$: Negation function making $C_1 = C_2 C_\nu$}        
        \For {$i = 1, 2, \ldots, n$} {
            $P_{in(i)} \gets [\ket{+}, \ket{+}, ..., \ket{+}]$
            
            $P_{in(i)}[i] \gets \ket{0}$

            $\nu(i) \gets 0$
            
            \For {$j = 1, 2, \ldots, k$} {
                \If {\Call{Swap\_Test}{$C_1(P_{in(i)}), C_2(P_{in(i)})$} = 1} {
                    $\nu(i) \gets 1$
                    
                    \textbf{break}
                }
            }  
        }
        
        \Return{$\nu$}     
\end{algorithm}

The correctness of \Cref{alg:input-negation} is established below.
We first show that $\nu(1)$ can be correctly decided, and $\nu(i)$ for $i=2,\ldots, n$ follow with the same derivation.
When deciding $\nu(1)$, the input states of $C_1$ and $C_2$ are both prepared as $\ket{\psi} = \ket{0} \otimes \ket{+} \otimes \ldots \otimes \ket{+}$.
The output states of $C_1$ and $C_2$ are $C_1 \ket{\psi}$ and $C_2 \ket{\psi}$, respectively.
Since $C_1 = C_2 C_\nu$, the output state of $C_1$ can also be written as $C_2 C_\nu \ket{\psi}$.
Let $C_\nu \ket{\psi}$ = $\ket{\psi'}$.
Note that a \textsc{not}-gate is a Pauli $X$ operator, whose application on $\ket{+}$ has no effect as $X\ket{+} = \ket{+}$. %
Hence, only the \textsc{not}-gate, if it exists, at the first bit in $C_\nu$ can make an effect on $\ket{\psi}$.
There are two cases:
In the first case, the first bit in $C_\nu$ has a \textsc{not}-gate, i.e., $\nu(1) = 1$.
Then $\ket{\psi'} = \ket{1} \otimes \ket{+} \otimes \ldots \otimes \ket{+}$, and
\(
        \langle \psi' | \psi \rangle
        = \langle 1 | 0 \rangle \langle + | + \rangle \cdots \langle + | + \rangle 
        = 0 \cdot 1 \cdots 1 = 0.
\)
Note that the final state of $C_1$ and $C_2$ are $C_2 \ket{\psi'}$ and $C_2\ket{\psi}$, respectively.
As mentioned in \Cref{sec:preliminaries}, quantum circuits must preserve the inner product of states.
Therefore, after applying $C_2$ on both $\ket{\psi'}$ and $\ket{\psi}$, the final states of the two circuits still have inner product $0$.
Finally, when applying the swap test on the output states of $C_1$ and $C_2$, there is a 50\% probability of obtaining a measurement outcome 1.

In the second case, the first bit in $C_N$ does not have a \textsc{not}-gate, i.e., $\nu(1) = 0$. 
In this case, the output states of $C_1$ and $C_2$ are identical.
When we apply the swap test to them, we always obtain a measurement outcome of 0.
Hence, once the measurement outcome is 1, we can conclude $\nu(1) = 1$.
Otherwise, if the measurement outcome is 0 for $k$ times, $\nu(1) = 0$ is of high confidence. 
The probability of success is $1 - 1/2^k$.
To make the failure probability $\leq \epsilon$, we require $k \geq \log_2(1/\epsilon)$.
Since a similar process is used to decide $\nu(i)$ for all $i = 1, \ldots, n$, the query complexity of \Cref{alg:input-negation} is $O(n\log(1/\epsilon))$.

\subsection{NP-I Equivalence}
\label{subsec:NP-I}
\begin{proposition}
For NP-I equivalence, finding $\nu$ and $\pi$ for $C_1 = C_2 C_\pi C_\nu$ is of $O(\log(n))$ query complexity if $C_1^{-1}$ or $C_2^{-1}$ is available, and of $O(n^2 \log(1/\epsilon))$ query complexity by a randomized quantum algorithm with success probability $1-\epsilon$ if $C_1^{-1}$ and $C_2^{-1}$ are unavailable but $C_1$ and $C_2$ can take quantum states as inputs.
\end{proposition}

If $C_1^{-1}$ or $C_2^{-1}$ is available, a method similar to that in \Cref{subsec:I-NP} can decide $\nu$ and $\pi$ in $O(\log(n))$ oracle queries.
If $C_1^{-1}$ and $C_2^{-1}$ are both unavailable, the following quantum algorithm can be applied.
First, find $C_\pi$ by deciding whether the $b_1^\mathrm{th}$ bit in $C_1$ is mapped to the $b_2^\mathrm{th}$ bit in $C_2$ for all $b_1, b_2 \in \{1, 2, \ldots, n\}$.
For each $(b_1, b_2)$ pair, initialize the $b_1^\mathrm{th}$ qubit in $C_1$ and the $b_2^\mathrm{th}$ qubit in $C_2$ to $\ket{-} =\frac{1}{\sqrt{2}}(\ket{0}-\ket{1})$, while all other qubits are initialized to $\ket{+}=\frac{1}{\sqrt{2}}(\ket{0}+\ket{1})$.
Then, execute both circuits and compare their final states by the swap test.
Let $\pi(b_2) = b_1$ if and only if the measurement outcome of the swap test is 0 for $k$ times.
After $\pi$ is found, \Cref{alg:input-negation} can be slightly modified to find $\nu$ further.

The key idea of the algorithm is to disable $C_\nu$ first, so as to focus on finding $C_\pi$.
To disable $C_\nu$, choose the input state to be $\ket{+}$ and $\ket{-}$.
Note that a \textsc{not}-gate, i.e., a Pauli $X$ operator, acting on $\ket{-}$ yields $X\ket{-} = 
-\ket{-}$, where the global phase $-1$ can be ignored.
Moreover, the input states of $C_1$ and $C_2$ are identical when $b_1 = b_2$ and are orthogonal when $b_1 \neq b_2$.
Therefore, the final states of $C_1$ and $C_2$ are identical if $\pi(b_2) = b_1$, and orthogonal otherwise.
By applying the swap test $k$ times on the final states of $C_1$ and $C_2$, we can decide whether $\pi(b_2) = b_1$.
Again, we can derive that $k = \log_2(1 / \epsilon)$ to make the failure probability $\leq \epsilon$, so the overall complexity of the algorithm is $O(n^2\log(1/\epsilon))$.

\subsection{P-N Equivalence}
\label{subsec:P-N}
\begin{proposition}
For P-N equivalence, finding $\pi$ and $\nu$ for $C_1 = C_\nu C_2 C_\pi$ is of $O(\log(n))$ query complexity if $C_1^{-1}$ or $C_2^{-1}$ is available, and of $O(n)$ query complexity if $C_1^{-1}$ and $C_2^{-1}$ are unavailable.
\end{proposition}
The result comes from the fact that this problem can be reduced to P-I equivalence testing, so it has the same complexity as P-I equivalence.
Regardless of whether $C_1^{-1}$ and $C_2^{-1}$ are available, we first decide $\nu$ by setting all inputs to 0 and observing the outputs.
Since all inputs are the same, input permutation makes no effect.
Then $\nu(i) = 1$ if and only if the output patterns of $C_1$ and $C_2$ differ at the $i^\mathrm{th}$ bit.
This step only takes $O(1)$ oracle queries and does not affect the complexity analysis.
After $\nu$ is found, we obtain $C_3 = C_\nu C_2$.
Note that $C_1$ and $C_3$ are P-I equivalent now, and the P-N equivalence can be reduced to P-I equivalence.

\subsection{N-P Equivalence}\label{subsec:N-P}
\begin{proposition}
For N-P equivalence, finding $\nu$ and $\pi$ for $C_1 = C_\pi C_2 C_\nu$ is of $O(\log(n))$ query complexity if $C_1^{-1}$ and $C_2^{-1}$ are both available.
\end{proposition}

If inverses of $C_1$ and $C_2$ are both available, the method mentioned in \Cref{subsec:P-N} is applicable to find $\nu$ and $\pi$ by the fact that $C_1^{-1} = C_{\nu^{-1}} C_2^{-1} C_{\pi^{-1}}$. %
Then $\nu^{-1}$ and $\pi^{-1}$ are transformed to $\nu$ and $\pi$.
The complexity is the same as P-N equivalence testing.

If $\nu$ and $\pi$ when both $C_1^{-1}$ and $C_2^{-1}$ are unavailable, whether there exists an efficient quantum algorithm remains an open problem for our future work.

\section{Intractable Equivalences}
\label{sec:discussions}

We show that the equivalence classes not solved in \Cref{sec:methods} are more difficult than the promise \emph{UNIQUE-SAT problem} \cite{promise}.
Given a conjunctive normal form (CNF) Boolean formula $\varphi$ that is promised to have at most one satisfying assignment, the UNIQUE-SAT problem asks to decide whether it is satisfiable.
It has been proved that SAT is randomly reducible to UNIQUE-SAT \cite{reduciable}, and thus UNIQUE-SAT is believed to be a difficult problem.
In what follows, we will show that the UNIQUE-SAT problem is polynomially reducible to N-N and P-P equivalences.
Since equivalences $\{$NP-NP, N-NP, NP-N, NP-P, P-NP$\}$ subsume N-N or P-P, they are thus harder than UNIQUE-SAT.

\subsection{Hardness of N-N Equivalence}
\label{subsec:IO-N}
\begin{theorem}
For N-N equivalence, finding $\nu_x$ and $\nu_y$ for $C_1 = C_{\nu_y} C_2 C_{\nu_x}$ is no easier than UNIQUE-SAT.
\end{theorem}
\vspace{-1em}
\begin{proof}
We establish a polynomial-time reduction from UNIQUE-SAT to N-N equivalence as follows.
Consider a CNF formula $\varphi = c_1 \land c_2 \land \ldots \land c_m$ over variables $\{x_1, x_2, \ldots, x_n\}$ promised to have at most one satisfying assignment.
Let clause $c_i = (\ell_{i,1} \lor \ell_{i,2} \lor \ldots \lor \ell_{i, k_i})$, a disjunction of $k_i$ literals. 
We construct the UNIQUE-SAT encoding reversible circuit $C_1$ as shown in the black part of \Cref{subfig:ION-overall} (excluding the red part), 
where the first $n$ bits $b_{x_1}, \ldots, b_{x_n}$ correspond to the input variables, 
the next $m$ ancilla bits $b_{a_1}, \ldots, b_{a_m}$ correspond to the valuations of the clauses, and 
two extra ancilla bits $b_b$ and $b_z$ are used to generate the value of $\varphi$.
Each $U(\varphi) = U(c_m) \cdots U(c_1)$ is the concatenation of the clause-encoding circuits. 
For a clause $c_i = \ell_{i,1} \lor \ldots \lor \ell_{i,k_i}$, its clause-encoding circuit is an MCT gate followed by a \textsc{not}-gate. 
In the MCT gate, each literal $\ell_{i,j}$ in $c_i$ corresponds to a control bit.
If $\ell_{i,j}$ equals $x_s$ (resp. $\overline{x_s}$), then a negative (resp. positive) control bit is applied on $b_{x_s}$.
The target bit of the MCT gate and the \textsc{not}-gate are applied on $b_{a_i}$.
E.g., the clause-encoding circuit of clause $c_1 = \overline{x_1} \lor x2 \lor \overline{x_3}$ is shown in \Cref{subfig:ION-clause}.
Hence, for the whole $U(\varphi)$ block, if the input value of $b_{x_j}$ equals $x_j$ and the input value of $b_{a_i}$ equals $a_i$, then the output values of all $b_{x_j}$ remain the same as their input values, while the output value of $b_{a_i}$ equals $a_i \oplus (\overline{\ell_{i,1}}\ldots\overline{\ell_{i,k_i}}) \oplus 1 = a_i \oplus c_i$.
Note that $U(\varphi)^{-1} = U(\varphi)$.

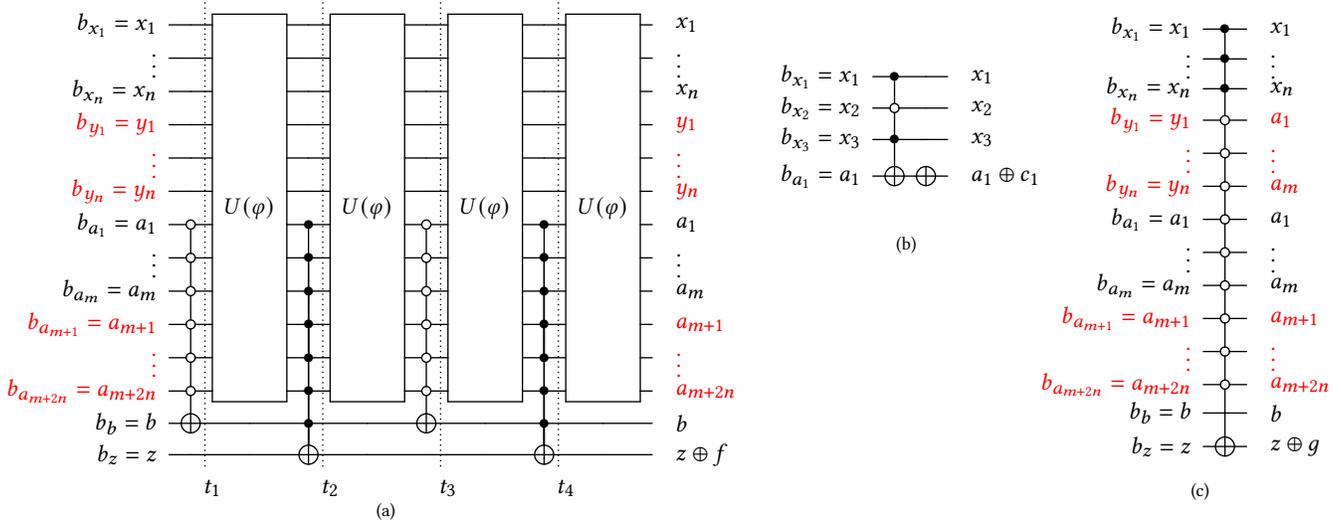
\begin{figure*}[t]
\centering
\subfigure[]{  \label{subfig:ION-overall} \label{subfig:IOP-overall}
    \Qcircuit @C=.5em @R=.5em { 
        &&&&\ar@{.}[]+<0.6em,0em>;[dddddddddddddd]+<0.6em,-0.5em>&&\ar@{.}[]+<0.6em,0em>;[dddddddddddddd]+<0.6em,-0.5em>&&\ar@{.}[]+<0.6em,0em>;[dddddddddddddd]+<0.6em,-0.5em>&&\ar@{.}[]+<0.6em,0em>;[dddddddddddddd]+<0.6em,-0.5em>                  \\
        &&&\lstick{b_{x_1} = x_1}           & \qw       & \multigate{11}{{U(\varphi)}} & \qw      & \multigate{11}{{U(\varphi)}} & \qw       & \multigate{11}{{U(\varphi)}} & \qw      & \multigate{11}{{U(\varphi)}} & \qw & \rstick{x_1}\\
        &&&\lstick{\vdots}                  & \qw       & \ghost{{U(\varphi)}}        & \qw      & \ghost{{U(\varphi)}}        & \qw       & \ghost{{U(\varphi)}}        & \qw      & \ghost{{U(\varphi)}}        & \qw & \rstick{\vdots}\\
        &&&\lstick{b_{x_n} = x_n}           & \qw       & \ghost{{U(\varphi)}}        & \qw      & \ghost{{U(\varphi)}}        & \qw       & \ghost{{U(\varphi)}}        & \qw      & \ghost{{U(\varphi)}}        & \qw & \rstick{x_n} \\
        &&&\lstick{\textcolor{red}{b_{y_1} = y_1}}           & \qw       & \ghost{{U(\varphi)}}        & \qw      & \ghost{{U(\varphi)}}        & \qw       & \ghost{{U(\varphi)}}        & \qw      & \ghost{{U(\varphi)}}        & \qw & \rstick{\textcolor{red}{y_1}}\\
        &&&\lstick{\textcolor{red}{\vdots}}                  & \qw       & \ghost{{U(\varphi)}}        & \qw      & \ghost{{U(\varphi)}}        & \qw       & \ghost{{U(\varphi)}}        & \qw      & \ghost{{U(\varphi)}}        & \qw & \rstick{\textcolor{red}{\vdots}}\\
        &&&\lstick{\textcolor{red}{b_{y_n} = y_n}}           & \qw       & \ghost{{U(\varphi)}}        & \qw      & \ghost{{U(\varphi)}}        & \qw       & \ghost{{U(\varphi)}}        & \qw      & \ghost{{U(\varphi)}}        & \qw & \rstick{\textcolor{red}{y_n}} \\
        &&&\lstick{b_{a_1} = a_1}           & \ctrlo{1} & \ghost{{U(\varphi)}}        & \ctrl{4} & \ghost{{U(\varphi)}}        & \ctrlo{1} & \ghost{{U(\varphi)}}        & \ctrl{4} & \ghost{{U(\varphi)}}        & \qw & \rstick{a_1}\\
        &&&\lstick{\vdots}                  & \ctrlo{1} & \ghost{{U(\varphi)}}        & \ctrl{3} & \ghost{{U(\varphi)}}        & \ctrlo{1} & \ghost{{U(\varphi)}}        & \ctrl{3} & \ghost{{U(\varphi)}}        & \qw & \rstick{\vdots}\\
        &&&\lstick{b_{a_{m}} = a_{m}} & \ctrlo{1} & \ghost{{U(\varphi)}}        & \ctrl{2} & \ghost{{U(\varphi)}}        & \ctrlo{1} & \ghost{{U(\varphi)}}        & \ctrl{2} & \ghost{{U(\varphi)}}        & \qw & \rstick{a_{m}} \\
        &&&\lstick{\textcolor{red}{b_{a_{m+1}} = a_{m+1}}}           & \ctrlo{1} & \ghost{{U(\varphi)}}        & \ctrl{4} & \ghost{{U(\varphi)}}        & \ctrlo{1} & \ghost{{U(\varphi)}}        & \ctrl{4} & \ghost{{U(\varphi)}}        & \qw & \rstick{\textcolor{red}{a_{m+1}}}\\
        &&&\lstick{\textcolor{red}{\vdots}}                  & \ctrlo{1} & \ghost{{U(\varphi)}}        & \ctrl{3} & \ghost{{U(\varphi)}}        & \ctrlo{1} & \ghost{{U(\varphi)}}        & \ctrl{3} & \ghost{{U(\varphi)}}        & \qw & \rstick{\textcolor{red}{\vdots}}\\
        &&&\lstick{\textcolor{red}{b_{a_{m+2n}} = a_{m+2n}}} & \ctrlo{1} & \ghost{{U(\varphi)}}        & \ctrl{2} & \ghost{{U(\varphi)}}        & \ctrlo{1} & \ghost{{U(\varphi)}}        & \ctrl{2} & \ghost{{U(\varphi)}}        & \qw & \rstick{\textcolor{red}{a_{m+2n}}} \\
        &&&\lstick{b_{b} = b}               & \targ     & \qw                      & \ctrl{1} & \qw                           & \targ     & \qw                      & \ctrl{1} & \qw                           & \qw & \rstick{b} \\
        &&&\lstick{b_{z} = z}               & \qw       & \qw                      & \targ    & \qw                           & \qw       & \qw                      & \targ    & \qw                           & \qw & \rstick{z \oplus f} \\\\
        &&&&\rlap{$\;\;t_1$}&&\rlap{$\;\;t_2$}&&\rlap{$\;\;t_3$}&&\rlap{$\;\;t_4$} \\\\
    }
}
\;\;\;\;\;\;\;\;\;\;\;\;\;\;\;\;\;\;\;\;\;\;\;\;\;\;\;\;
\subfigure[]{ \label{subfig:ION-clause}
    \centering
    \Qcircuit @C=.5em @R=1em { 
        \\\\\\
        \lstick{b_{x_1} = x_1} & \ctrl{1}  & \qw   & \qw   & \rstick{x_1}\\
        \lstick{b_{x_2} = x_2} & \ctrlo{1} & \qw   & \qw   & \rstick{x_2}\\
        \lstick{b_{x_3} = x_3} & \ctrl{1}  & \qw   & \qw   & \rstick{x_3}\\
        \lstick{b_{a_1} = a_1} & \targ     & \targ & \qw   & \rstick{a_1 \oplus c_1} \\\\\\
    }
}
\;\;\;\;\;\;\;\;\;\;\;\;\;\;\;\;\;\;\;\;\;\;\;\;\;\;\;
\subfigure[]{   \label{subfig:ION-dual} \label{subfig:IOP-dual}
    \Qcircuit @C=.5em @R=1em { \\
        &&&\lstick{b_{x_1} = x_1}           & \ctrl{1}  & \qw & \rstick{x_1}\\
        &&&\lstick{\vdots}                  & \ctrl{1}  & \qw & \rstick{\vdots}\\
        &&&\lstick{b_{x_n} = x_n}           & \ctrl{1}  & \qw & \rstick{x_n} \\
        &&&\lstick{\textcolor{red}{b_{y_1} = y_1}}           & \ctrlo{1} & \qw & \rstick{\textcolor{red}{a_1}}\\
        &&&\lstick{\textcolor{red}{\vdots}}                  & \ctrlo{1} & \qw & \rstick{\textcolor{red}{\vdots}}\\
        &&&\lstick{\textcolor{red}{b_{y_n} = y_n}}           & \ctrlo{1} & \qw & \rstick{\textcolor{red}{a_m}} \\
        &&&\lstick{b_{a_1} = a_1}           & \ctrlo{1} & \qw & \rstick{a_1}\\
        &&&\lstick{\vdots}                  & \ctrlo{1} & \qw & \rstick{\vdots}\\
        &&&\lstick{b_{a_m} = a_m}           & \ctrlo{1} & \qw & \rstick{a_m} \\
        &&&\lstick{\textcolor{red}{b_{a_{m+1}} = a_{m+1}}}   & \ctrlo{1} & \qw & \rstick{\textcolor{red}{a_{m+1}}}\\
        &&&\lstick{\textcolor{red}{\vdots}}                  & \ctrlo{1} & \qw & \rstick{\textcolor{red}{\vdots}}\\
        &&&\lstick{\textcolor{red}{b_{a_{m+2n}} = a_{m+2n}}} & \ctrlo{2} & \qw & \rstick{\textcolor{red}{a_{m+2n}}} \\
        &&&\lstick{b_{b} = b}               & \qw       & \qw & \rstick{b} \\
        &&&\lstick{b_{z} = z}               & \targ     & \qw & \rstick{z \oplus g} \\\\
    }
}   
\caption{
(a) The UNIQUE-SAT encoding circuit, (b) a clause-encoding circuit $U(c)$ of clause $c = \overline{x_1} \lor x2 \lor \overline{x_3}$, (c) the $C_2$ circuit for Boolean matching. 
}
\label{fig:ION-SAT-encode}
\end{figure*}

For the overall UNIQUE-SAT encoding circuit in \Cref{subfig:ION-overall}, let $v$ denote the input value of bit $b_v$.
We note that the value of $b_{a_i}$ is changed to $a_i \oplus c_i$ at time $t_2$ and $t_4$ and is returned back to $a_i$ at time $t_2$ and the end of the circuit.
Moreover, the value of $b_b$ is changed to $b \oplus (\overline{a_1}\ldots\overline{a_m})$ at time $t_1$ and returned back to $b$ at time $t_3$.
It can be derived that the output values of the first $n+m+1$ bits are the same as their input values, while the output value of $b_z$ is $z \oplus f$, for 
\begin{equation}
    \begin{aligned}
        f &= \left[ \bigwedge_{i=1}^m (a_i \oplus c_i) (b \oplus (\overline{a_1}\ldots\overline{a_m}))\right]
          \oplus \left[\bigwedge_{i=1}^m (a_i \oplus c_i)  b\right] \\
          &= [(a_1 \oplus c_1)\ldots(a_m \oplus c_m)]\land (\overline{a_1}\ldots\overline{a_m}) \\
          &= (c_1 \ldots c_m) \land (\overline{a_1} \ldots \overline{a_m})
          \;\;\;= \varphi \land (\overline{a_1} \ldots \overline{a_m}).
    \end{aligned}
\end{equation}
Hence, $f = 1$ if and only if all $a_i$'s are 0 and $(x_1, \ldots, x_n)$ is a satisfying assignment of $\varphi$.

On the other hand, we can construct a circuit $C_2$ shown in \Cref{subfig:ION-dual}. 
Let $v$ denote the input value of bit $b_v$.
It is easy to derive that the output values of the first $n+m+1$ bits are the same as their input values, while the output value of $b_z$ is $z \oplus g$, where $g = (x_1 \ldots x_n)\land(\overline{a_1} \ldots \overline{a_m})$.
Taking the UNIQUE-SAT encoding circuit as $C_1$ and comparing $f$ and $g$,
it can be verified that $C_1$ and $C_2$ are N-N equivalent if and only if $\varphi$ is satisfiable.
The intuition is as follows.
First, $\pi_x(i) = \pi_y(i)$ for $i = 1, \ldots, n+m+1$ must hold to ensure the output values of the first $n+m+1$ bits are the same as their input values. 
Then we observe that two \textsc{not}-gates applied before and after a control bit will flip the control polarity. Therefore, if $v_1(i) = v_2(i) = 1$, then $x_i$ is in fact negatively controlling, indicating that $x_i = 0$ is the necessary condition to make $\varphi = 1$.
Otherwise, $v_1(i) = v_2(i) = 0$ indicates that $x_i$ is positively controlling, so $x_i = 1$ is the necessary condition to make $\varphi = 1$.
Therefore, once we can find $\nu_x$ and $\nu_y$ to make $C_1$ and $C_2$ N-N equivalent, the unique satisfying assignment of $\varphi$ is found.
Even if we do not know whether $C_1 = C_{\nu_y} C_2 C_{\nu_x}$, we can still try the process and obtain a candidate solution.
The validity of the candidate solution can be easily verified in linear time by substituting it into $\varphi$.
Moreover, the reduction process is polynomial since there are only $8m + 4$ MCT gates in the UNIQUE-SAT encoding circuit.
Hence, UNIQUE-SAT is polynomially reducible to the N-N equivalence problem, and the theorem follows.
\end{proof}

\subsection{Hardness of P-P Equivalence}
\label{subsec:IO-P}
\begin{theorem}
For P-P equivalence, finding $\pi_x$ and $\pi_y$ for $C_1 = C_{\pi_y} C_2 C_{\pi_x} $ is no easier than UNIQUE-SAT.
\end{theorem}
\vspace{-1em}
\begin{proof}
Consider again the CNF formula $\varphi$ in \Cref{subsec:IO-N}.
We create another CNF formula $\varphi'$ by adding $n$ extra variables $y_1, \ldots, y_n$ and setting $\varphi' = \varphi \land c_{m+1} \land \ldots \land c_{m+2n}$, where $(c_{m+2j - 1}) \land (c_{m+2j}) = (x_j \lor y_j) \land (\overline{x_j} \lor \overline{y_j}), j = 1, \ldots n$.
That is, we make a dual-rail encoding on the variables of $\varphi$ in $\varphi'$ and set $y_j = \overline{x_j}$ for all $y_j$.
Hence, $\varphi$ is satisfiable if and only if $\varphi'$ is satisfiable.
Then the same method mentioned in $\Cref{subsec:IO-N}$ is applied to encode $\varphi'$ into $C_1$,
as shown in \Cref{subfig:IOP-overall} (including both the black and red parts).
Again, the output values of the first $4n+m+1$ bits are always the same as their input values.
For the last bit $b_z$, if its input value is $z$, then its output value is $z \oplus f$, where $f = \varphi' \land (\overline{a_1} \ldots \overline{a_{m+2n}})$.
On the other hand, we can construct a circuit $C_2$ shown in \Cref{subfig:IOP-dual} (including both black and red parts), where the first $n$ bits are positive-control bits, the ${(n+1)}^\text{th}$ to the ${(4n+m)}^\text{th}$ bits are negative-control bits, and the last bit $b_z$ is the target bit.

By comparing $f$ and $g$,
it can be verified that $C_1$ and $C_2$ are P-P equivalent if and only if $\varphi'$ is satisfiable.
The intuition is as follows.
First, we note that $\pi_x^{-1} = \pi_y$ must hold to ensure the output values of the first $4n + m + 1$ bits are the same as their input values.
Therefore, the input and output permutations are equivalently just permuting the control bits.
Second, the $i^\text{th}$ bit is positively controlling if it falls in the positive-control region ($0 < \pi_x^{-1}(i) \leq n$), or it is negatively controlling if it falls in the negative-control region ($n < \pi_x^{-1}(i) \leq 4n+m$).
If $x_i$ is permuted to the positive-control region and $y_i$ is permuted to the negative-control region, then $x_i = \overline{y_i} = 1$ is the necessary condition to make $\varphi' = 1$.
Otherwise, if $x_i$ is permuted to the negative-control region and $y_i$ is permuted to the positive-control region, then $x_i = \overline{y_i} = 0$ is the necessary condition to make $\varphi' = 1$.
Therefore, once we can find $\pi_x$ and $\pi_y$ to make $C_1$ and $C_2$ P-P equivalent, the unique satisfying assignment of $\varphi'$ is found, which can be easily transformed to the unique satisfying assignment of $\varphi$.
Even if we do not know whether $C_1$ and $C_2$ are P-P equivalent or not, we can still try the process and obtain a candidate solution.
The validity of the candidate solution can be easily verified by substituting it into $\varphi$.
Moreover, the reduction process is polynomial since there are only $8m + 4$ MCT gates in the UNIQUE-SAT encoding circuit.
Hence, UNIQUE-SAT is polynomially reducible to the P-P equivalence problem, and the theorem follows.
\end{proof}

\section{Conclusions and Future Work}
\label{sec:conclusions}

This work provided the first comprehensive study on various equivalences for Boolean matching of reversible circuits by characterizing their computational complexities. 
For the tractable equivalences, polynomial-time (classical or quantum) algorithms were devised.
For the intractable equivalences, their hardness results were established.
The foundation paved in this work may open new Boolean matching applications, e.g., in template-based reversible logic synthesis and in quantum program compilation for oracle circuit minimization.
Moreover, our swap-test-based algorithm demonstrates the first example with an exponential quantum speedup over classical computation in design automation research.
It may inspire the development of new types of quantum algorithms and applications.
For future work, we intend to resolve the remaining open problem regarding the quantum complexity of N-P equivalence.

\section*{Acknowledgments}
This work was supported in part by the National Science and Technology Council of Taiwan under grants 112-2119-M-002-017 and 113-2119-M-002-024, and the NTU Center of Data Intelligence: Technologies, Applications, and Systems under grant NTU-113L900903.
The authors thank IBM Q Hub at NTU and Quantum Technology Cloud Computing Center at NCKU for supporting experimental validation.

\bibliographystyle{ACM-Reference-Format}
\bibliography{reference}

\end{document}